\documentclass[sigconf]{acmart}
\AtBeginDocument{%
  }

\copyrightyear{2025}
\acmYear{2025}
\setcopyright{cc}
\setcctype{by}
\acmConference[CHI '25]{CHI Conference on Human Factors in Computing Systems}{April 26-May 1, 2025}{Yokohama, Japan}
\acmBooktitle{CHI Conference on Human Factors in Computing Systems (CHI '25), April 26-May 1, 2025, Yokohama, Japan}
\acmDOI{10.1145/3706598.3713292}
\acmISBN{979-8-4007-1394-1/25/04}

\newcommand{\minititle}[1]{\noindent
\textbf{#1}}

\usepackage{hyperref}
\usepackage{multirow}
\usepackage{listings}
\usepackage{array}
\usepackage{sparklines}

\begin{document}

\title{AI Rules? Characterizing Reddit Community Policies Towards AI-Generated Content}

\author{Travis Lloyd}
\email{tgl33@cornell.edu}
\orcid{0009-0009-7393-7105}
\affiliation{%
  \institution{Cornell Tech}
  \city{New York}
  \state{NY}
  \country{USA}
}

\author{Jennah Gosciak}
\email{jrg377@cornell.edu}
\orcid{0009-0007-2180-8147}
\affiliation{%
  \institution{Cornell University}
  \city{Ithaca}
  \state{NY}
  \country{USA}
}

\author{Tung Nguyen}
\email{tn375@cornell.edu}
\orcid{0009-0006-6456-4931}
\affiliation{%
  \institution{Cornell Tech}
  \city{New York}
  \state{NY}
  \country{USA}
}

\author{Mor Naaman}
\email{mor.naaman@cornell.edu}
\orcid{0000-0002-6436-3877}
\affiliation{%
  \institution{Cornell Tech}
  \city{New York}
  \state{NY}
  \country{USA}
}

\renewcommand{\shortauthors}{Lloyd et al.}

\begin{abstract}
  How are Reddit communities responding to AI-generated content? 
  We explored this question through a large-scale analysis of subreddit community rules and their change over time. 
  We collected the metadata and community rules for over $300,000$ public subreddits and measured the prevalence of rules governing AI.
  We labeled subreddits and AI rules according to existing taxonomies from the HCI literature and a new taxonomy we developed specific to AI rules.
  While rules about AI are still relatively uncommon, the number of subreddits with these rules more than doubled over the course of a year.
  AI rules are more common in larger subreddits and communities focused on art or celebrity topics, and less common in those focused on social support.
  These rules often focus on AI images and evoke, as justification, concerns about quality and authenticity.
  Overall, our findings illustrate the emergence of varied concerns about AI, in different community contexts.
  Platform designers and HCI researchers should heed these concerns if they hope to encourage community self-determination in the age of generative AI.
  We make our datasets public to enable future large-scale studies of community self-governance.
\end{abstract}

\ccsdesc[500]{Human-centered computing~Empirical studies in collaborative and social computing}

\keywords{Online Communities, Reddit, Rules, Generative AI, Governance, AI-Generated Content, Moderation}

\received{12 September 2024}
\received[revised]{11 December 2024}
\received[accepted]{27 January 2025}

\maketitle

\section{Introduction}

The emergence of mainstream generative AI tools capable of producing compelling text, images, and videos---collectively referred to as AI-generated content (AIGC)---poses a unique challenge for online platforms that are dependent on user-generated content. 
Many such platforms have enacted top-down changes to respond to this new technology, with some banning it~\cite{makyen23}, while others are encouraging its use or incorporating generative AI features into their user interfaces~\cite{lundenLinkedInNow1B2023a,mehtaInstagramIntroducesGenAI2023}.
Other platforms have declined to make top-down changes and have instead left it up to users to come to their own positions on this technology.
One such platform is Reddit, a social sharing and news aggregation site, where individual communities, known as subreddits, are free to enact their own rules and policies~\cite{fiesler2018reddit,proferesStudyingRedditSystematic2021a,singer14,gilber13}.
This decentralized moderation approach allows communities to craft generative AI policies that are contextually-relevant: the policies are based on how communities observe the technology being used, and whether or not that use aligns with their values. 
Decentralized moderation also provides a rich source of data through which HCI researchers interested in helping online communities navigate sudden technological change can better understand the range of attitudes and stances that exist across communities who vary in size, purpose, and values.

Findings from past HCI studies have emphasized the important role that explicit and consistent community rules play in helping online communities navigate moments of intense change~\cite{kieneSurvivingEternalSeptember2016a}.
These rules interact with other forms of governance~\cite{lessigCodeOtherLaws2009}, such as norms~\cite{Chandrasekharan18} and technical affordances~\cite{Jhaver19}, to encourage certain behaviors and discourage others~\cite{kieslerRegulatingBehaviorOnline2012}. 

Of course, generative AI and AIGC may impact the rules~\cite{makyen23}, dynamics~\cite{hua24,fiesler_ai_2024}, and governance~\cite{caoToxicityDetectionNOT2023a} strategies of online communities in complex ways.
While some recent research has explored how generative AI might help online communities deal with problematic content~\cite{agarwalConversationalAgentsFacilitate2024,kumarWatchYourLanguage2024}, other work has documented the ability of AIGC to pollute these same communities with deceptive information~\cite{park_ai_2024}.
In our own earlier work~\cite{lloyd_there_2023}, we used interviews with Reddit moderators to explore the particular challenges of governing AI use in online communities and found that communities are enacting explicit rules about AIGC in order to clarify norms.
This earlier study focused on the perspectives and experiences of the interviewees and raised the question of how the findings generalize across the wide variety of communities on Reddit.

This paper expands our understanding of Reddit communities' responses to AIGC through a large-scale quantitative analysis of community rules.
Our approach allows us to paint a fuller picture for the HCI community by complementing the earlier qualitative findings with an analysis of platform-wide patterns and trends, while simultaneously providing deeper insight into the specific types of AI rules that communities are creating.
We build on techniques developed by~\citet{fiesler2018reddit} to study the ecosystem of community rules on Reddit and adapt their methodology to specifically investigate the emergence of community rules governing AI use.
We use Fiesler et al.'s taxonomies for community topics and rules so that we can compare our analysis of AI rules to their analysis of general Reddit rules.
We address the following research questions:
\begin{itemize}
    \item \textbf{RQ1:} How common are subreddit rules governing the use of AI and how has their prevalence changed over time?
    \item \textbf{RQ2:} What types of communities are more likely to have these rules?
    \item \textbf{RQ3:} What types of rules exist for governing the use of AI?
\end{itemize}

To answer these questions we crawled over $300,000$ public subreddits on two occasions, in July~2023 and November~2024, building a dataset of $N=99,969$ English-language subreddits with published rules.
Analyzing a longitudinal sample of subreddits that were present during both our crawls, our inquiry into \textbf{RQ1} finds that while in November~2024 only $1.2\%$ of subreddits had rules about AI, the number of subreddits with AI rules more than doubled since July~2023.
Larger subreddits are much more likely to have AI rules: for example, of the top $1\%$ largest subreddits in our 2024 sample, $17\%$ had AI rules.
To explore \textbf{RQ2}, we perform subreddit-level analysis on all subreddits with rules in our 2024 sample.
We apply a ``Topic'' label to these subreddits using a taxonomy from Fiesler et al.~\cite{fiesler2018reddit} and ``Community Archetype'' labels using a taxonomy from Prinster et al.~\cite{prinster2024community}. 
We use quantitative methods to explore the relationship between these labels and the presence of AI rules and find these rules to be most common in communities with \textit{Art} and \textit{Celebrity} topics, as well as communities that fit the \textit{Content Generation} Community Archetype. 
To explore \textbf{RQ3}, we perform rule-level analysis on the rules in our sample that we identified as governing AI use. 
We apply an existing rule taxonomy from Fiesler et al.~\cite{fiesler2018reddit} as well as a novel \textit{AI Rules Taxonomy} to describe meaningful dimensions of variation in both the requirements imposed by these rules and the language through which they frame AI usage. 
We find that rules about AI images are more common than rules about other types of AIGC, and see quality and authenticity as the most frequently mentioned concerns. 

Our findings offer implications for designers of online community platforms.
In particular, community concerns about quality and authenticity may be addressed using thoughtful design choices that encourage \textit{effortful communication}~\cite{Kelly17,Zhang22}.
Our findings also show, though, that communities of different types take varied approaches towards regulating AI use, and suggest that design solutions meant to help communities address these challenges must be sensitive to community context.

As an additional contribution, we make our community rules data and labels public\footnote{\url{https://github.com/sTechLab/AIRules}}. 
While we limit our current analysis to questions about AI rules, our data can be used to answer a wide range of questions about governance in self-moderating online communities. 
We hope that our data will enable large-scale analysis by future HCI researchers interested in encouraging online community self-determination.

\section{Related Work}
Our study is motivated by two bodies of work from the HCI literature.
The first includes empirical research into online communities' rules, norms, and values.
This work demonstrates that online communities hold a wide range of values and emphasizes that rules and norms are context-dependent.
The second body of work concerns current and potential future impacts of generative AI on online communities.
This research shows that communities are already grappling with the changes brought by generative AI and are adapting their policies and practices in response.

\subsection{Online Communities' Rules, Norms, and Values}
Online communities play an important role in social life~\cite{preeceEmpathicCommunitiesReaching1998}, and occupy a complicated position as a site of nuanced interactions between people, technology, and policy~\cite{jackson_policy_2014}.
This importance and complexity have made online communities a frequent topic of study in the HCI literature~\cite{preece2000online,krautBuildingSuccessfulOnline2012a}.
HCI scholars have explored the variety of purposes that these communities can serve and how platform design choices may influence their success.
For example, Preece introduced the idea of \emph{Empathic Communities}~\cite{preeceEmpathicCommunitiesBalancing1999a}, in which people come together to share and receive support, and suggested platform design guidelines to encourage sociality~\cite{preeceSociabilityUsabilityOnline2001}. 
In addition to encouraging pro-social behavior, a frequent topic of study has been how to discourage anti-social behavior~\cite{kieslerRegulatingBehaviorOnline2012,lessigCodeOtherLaws2009}.
Much of this work has focused on the benefits~\cite{grimmelmannVirtuesModeration2015a} and challenges~\cite{Koshy23} of effective and fair content moderation.
The HCI community has been interested in top-down moderation~\cite{gillespieCustodiansInternetPlatforms2018}, in which platforms set and enforce rules, as well as bottom-up, or self-moderation, in which communities perform these tasks~\cite{seeringSelf}.
A major result of this body of research is a recognition of the importance of context sensitivity in moderation decisions, which self-moderation is often more attuned to.
Taken together, this research suggests that studying self-moderation decisions may provide insight for design solutions that can promote community self-determination in times of sudden technological change.

Reddit is a useful site for a study of community self-moderation because it is home to online communities of various size and purpose.
This heterogeneity, combined with its largely public and text-based nature, have made Reddit a frequent topic of study in online communities research~\cite{proferesStudyingRedditSystematic2021a}.
Recent work has used surveys and digital trace data to explore the values~\cite{weldMakingOnlineCommunities2024}, norms~\cite{Chandrasekharan18}, and rules~\cite{fiesler2018reddit} of different communities across the platform.
Reddit's implementation of community self-moderation allows communities to write and make public their own community-enforced rules, which Fiesler et al. collect and characterize in a study that is the main methodological inspiration for our work~\cite{fiesler2018reddit}.
In their study, Fiesler et al. find subreddit size to be the main predictor of having rules, but note that even among subreddits of similar sizes rules are context-dependent, not uniform, and often uniquely tailored to the topic of the subreddit (e.g., art subreddits have more rules about copyright). 
The authors also highlight that setting norms and developing rules is fundamentally a social process.
Overall, Fiesler et al. show the importance of community rules and the feasibility of using them as a data source to study attitudes and stances across a platform.
Given the potential impact of AI on online communities, we use community rules as a starting point to study context-specific attitudes towards AIGC as they emerge across Reddit.

\subsection{Generative AI's Impact on Online Communities}
A growing body of research in HCI, Social Computing, and adjacent behavioral science fields has documented ways in which generative AI may impact online communities by increasing deception, decreasing trust, and enabling new forms of harassment.
Studies have demonstrated that AI can effectively deceive humans~\cite{park_ai_2024} and have raised concerns that generated text could be weaponized for deception by either humans~\cite{goldstein23} or automated social media accounts (bots)~\cite{menczer_addressing_2023}.
Additionally, just the \emph{suspicion} that other online community members are using AI may be enough to impact dynamics in online communities. 
Several studies in AI-Mediated Communication (AIMC)~\cite{hancock_aimc_2020} provide evidence that the perception that another online community member is using AI causes members to regard each other more negatively~\cite{rae_effects_2024} and trust each other less~\cite{Jakesch19}.
Finally, HCI research has long documented how online communities can be sites of harassment and abuse~\cite{jiang19,park22,cheng17,han_hate_2023}.
The impact of AI-powered harassment on online communities, such as malicious synthetic, or deepfake, media~\cite{farid_creating_2022} is much less studied~\cite{ncsii24}, but there is concern that this extremely harmful phenomenon may grow in prevalence alongside the popularity of generative AI tools~\cite{rini_deepfakes_2022}.

Many additional studies have started to document the impacts of AIGC on online communities ``in the wild''.
Recent work has identified undisclosed AI use in several settings, such as LLM-powered social bots on Twitter~\cite{yangAnatomyAIpoweredMalicious2024}, AIGC in fake online reviews~\cite{Oak24}, and in spam content used for scams~\cite{diresta2024spammers}.
In addition to studies of deceptive use, other recent work has begun to explore how the disclosed use of AIGC may affect community dynamics, such as in art-sharing communities~\cite{wei2024,matatov2024examiningprevalencedynamicsaigenerated}.
Generative AI can additionally impact online communities by drawing traffic away and decreasing overall engagement: studies of Q\&A sites in the Stack Exchange network, where users gather to ask and answer questions about programming and related topics, have demonstrated that the launch of ChatGPT caused an overall decline in website visits and question volume~\cite{burtch_consequences_2024}.
The authors of this study contrast this decline with a measured null-effect on similar Reddit communities, which they suggest may avoid a decline in traffic because of their emphasis on socialization, rather than pure information exchange. 
Finally, our previous work used interviews to qualitatively explore the attitudes and experiences of Reddit moderators dealing with AIGC~\cite{lloyd_there_2023}.
In that study, we interviewed moderators from subreddits of different sizes and topics.
This earlier work surfaces a variety of community objections to AIGC that range from practical to ideological. 
The findings of this qualitative study pose open questions about how prevalent these concerns are across the platform and suggest that these trends could be explored quantitatively in future work.
This current study fills that gap through a quantitative analysis that helps identify the most common rules and the most impacted communities, so that the HCI community can begin the task of addressing these issues with context-specific solutions.

\section{Methods}
\begin{figure*}[t]
    \centering
    \includegraphics[width=1\linewidth]{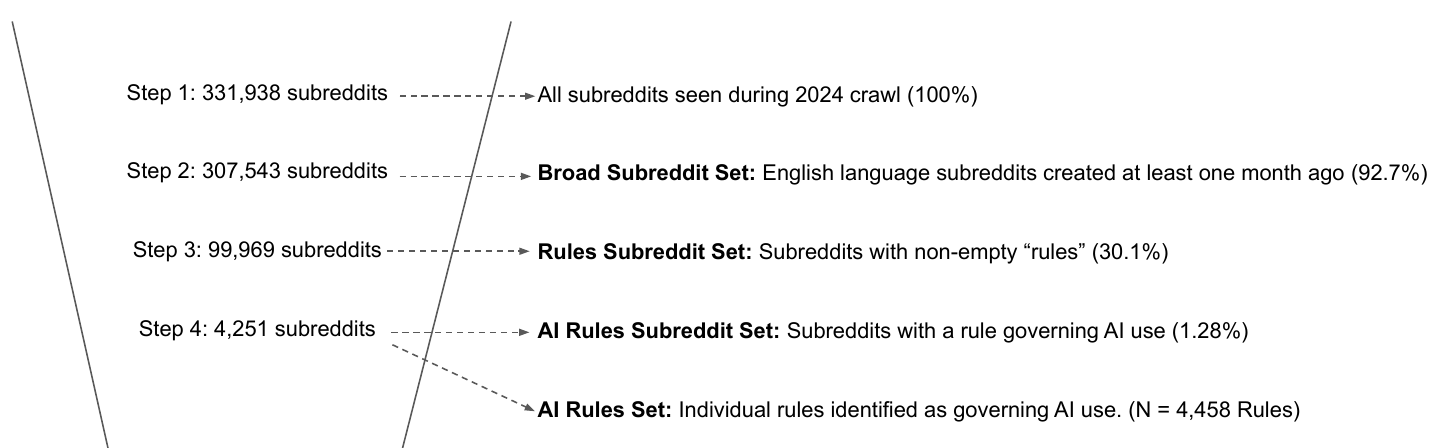}
     \caption{The process for collecting subreddit metadata and building the datasets used in our analysis.}
    \label{fig:data-funnel}
    \Description{A funnel diagram with four steps showing the filtering of subreddits based on various criteria:
1. Step 1: 331,938 subreddits, representing 100\% of all subreddits seen during the 2024 crawl.
2. Step 2: 307,543 subreddits (92.7\%), filtered to subreddits older than 1 month old where English is the primary language.
3. Step 3: 99,969 subreddits (30.1\%), filtered to subreddits with non-empty "rules".
4. Step 4: 4,251 subreddits (1.28\%), further filtered to subreddits with rules governing AI use.
The final output is 4,458 individual AI rules identified in these subreddits.}
\end{figure*}

In order to perform a large-scale analysis of community rules governing the use of AI, we collected the metadata of public, English-language subreddits.
Building on a large-scale crawl in November 2024, and an earlier one performed in July 2023, we created five datasets of subreddits and their associated rules.
We processed the more recent crawl to create four datasets: the \textbf{Broad Subreddit Set} ($N=307,543$) consisting of the metadata of subreddits with and without rules; the \textbf{Rules Subreddit Set} ($N=99,969$) of only subreddits with stated, published rules; the \textbf{AI Rules Subreddit Set} ($N=4,251$) of subreddits with rules governing the use of AI; and the \textbf{AI Rules Set} ($N=4,458$) of individual rules governing AI use.
The process of constructing these datasets is summarized in Figure~\ref{fig:data-funnel}, and detailed below. 
We supplement the raw crawl data with additional information about subreddit and rule types through several rounds of labeling, using both human and LLM coders.
We use emergent coding to construct a novel taxonomy of AI Rule Types from the \textbf{AI Rules Set}.
In addition, we created a longitudinal dataset of subreddits that were present in both the~2023 and 2024~crawls.
We use this \textbf{Longitudinal Subreddit Set} ($N=227,737$) to compare the prevalence of AI rules at different points in time for a consistent sample of subreddits.
Our datasets and labels are publicly available at \url{https://github.com/sTechLab/AIRules}.

\subsection{The Broad Subreddit and Rules Subreddit Sets}
In order to better understand how online communities across Reddit are responding to the arrival of AIGC, we conducted a large-scale data collection and analysis.
In November 2024 we performed an automated crawl of Reddit's ``Top Communities'' list\footnote{https://www.reddit.com/best/communities/1/}, which includes public subreddits with more than ten subscribers.
We used a web-scraping library\footnote{https://docs.scrapy.org/en/latest/} to paginate through this list and extract the IDs of all listed subreddits from each page.
We then used Reddit's Data API\footnote{https://support.reddithelp.com/hc/en-us/articles/16160319875092-Reddit-Data-API-Wiki} to fetch metadata and rules for each subreddit ID.

From the $N=331,938$ subreddits seen in our crawl, we used a process summarized in Figure~\ref{fig:data-funnel} to create several datasets for analysis.
We used an automated tool\footnote{https://pypi.org/project/langdetect/} to determine the language of each subreddit based on the text in its description and rules metadata.
We removed subreddits with a primary language other than English or a creation date less than one month before the crawl date, in order to remove communities that had not yet had time to create rules.
This process yielded the \textbf{Broad Subreddit Set} of $N=307,543$ public, English-language subreddits that were at least one month old, which we used to understand platform-wide patterns in the data and factors that contribute to whether subreddits have rules at all.

Using this broad dataset, we constructed a second, smaller dataset: the \textbf{Rules Subreddit Set} ($N=99,969$).
This dataset consists only of subreddits with non-empty \verb|rules| metadata. 
We use this dataset to understand patterns specific to subreddits with rules. 
These two datasets are the focus of our descriptive analysis and were used to create two other datasets specific to \textit{AI rules}, which are discussed in Sections \ref{emergent-rule-types} and \ref{sub-labels}.

\subsection{Longitudinal Subreddit Set}
To address RQ1's concern with the shift in rule prevalence over time, we constructed a \textbf{Longitudinal Subreddit Set}, using a July 2023 data collection effort as a baseline, and identifying the subset of subreddits from the 2024 crawl that were seen in the earlier crawl. 
ChatGPT launched at the end of 2022, kicking off a period of increased public awareness and usage of AI. 
Given the dynamic nature of such a moment of flux, a longitudinal analysis could provide  insight into the evolution of community norms and practices in a significant period soon after this launch.

The 2023 crawl used the same technique as the 2024 crawl, capturing subreddits in Reddit's ``Top Communities'' list as described above. 
In the 2023 crawl, we scraped the subreddit metadata and rules data directly from the web page of each subreddit, using the same tool that we used to visit the ``Top Communities'' list---a simpler approach that was possible under Reddit's prior rate limit.
We verified that the data that we gathered from a subreddit's web page is identical to the data returned by the API, which we used for the 2024 crawl.
This 2023 crawl scraped $N=337,399$ subreddits in total.
From this 2023 dataset, we identified the subreddits whose primary language was English, and that were also present in the 2024 crawl.
This process yielded the \textbf{Longitudinal Subreddit Set} ($N=227,737$) which we used to study the change in prevalence of AI rules over time.

\subsection{Data Labeling}
To answer our research questions about rules governing the use of AI, we created and applied labels to both the communities with rules and the AI rules contained in our sample. 
Through the process described below, we used a hybrid human-LLM coding technique to determine whether each of the subreddits in the \textbf{Rules Subreddit Set} and the \textbf{Longitudinal Subreddit Set} had an explicit rule about AI use, and to label them accordingly.
We used these labels to identify the subset of subreddits with AI rules.
We refer to the subset of the \textbf{Rules Subreddit Set} with AI rules as the \textbf{AI Rules Subreddit Set} ($N=4,251$).
The specific AI-related rules from this set of subreddits became our \textbf{AI Rules Set} ($N=4,458$).
We used an iterative inductive coding technique to identify common categories that emerged from the AI rules, and a hybrid human-LLM coding technique to label the rules according to these categories, which we describe in detail below.

Section~\ref{detection} details our process for detecting AI rules. 
Next, Section~\ref{emergent-rule-types} describes how we identified categories of AI rules.
Section~\ref{rule-class-labels} then describes how we labeled our data with these categories, and then finally, Section~\ref{sub-labels} describes how we applied community type labels from several existing frameworks to all subreddits in the \textbf{Rules Subreddit Set}.

\begin{table*}
\centering
\caption{Labels applied to rules in the \textbf{AI Rules Set}}
\label{tab:ruletype-labels}
\small 
\begin{tabular}{>{\raggedright}m{2.1cm}>{\raggedright}m{1.7cm}>{\raggedright\arraybackslash}m{2.6cm}>{\raggedright\arraybackslash}m{6.7cm}}
\hline
\textbf{Label Category} & \textbf{Source} & \textbf{Mutually Exclusive?} & \textbf{Example Labels} \\ \hline
Rule Type & Fiesler et al. & No & Advertising \& Commercialization, Prescriptive, Restrictive \\ 
AI Rule: Rationale & New & No & Low Quality, Inaccurate, Inhuman \\ 
AI Rule: Stance & New & No & Unqualified Ban, Qualified Ban, Disclosure Required \\ 
AI Rule: Medium & New & No & Text, Images, Deepfakes \\ 
AI Rule: Framing & New & No & Generate/Create, Edit/Assist/Enhance, Unspecified \\ \hline
\end{tabular}
\end{table*}

\subsubsection{Detecting Community Rules Governing AI Use}\label{detection}
Community rules are often long and complex, which made our datasets too large to manually code for the presence of AI rules.
To get a rough sense of the prevalence of these rules, we first used a simple regex to search the subreddits' rules for AI-related keywords.
This step found matches in the rules of less than $2\%$ of the subreddits.
We then devised a protocol, inspired by recent HCI work using LLMs for qualitative labeling tasks~\cite{choksi_under_2024,rao_quallm_2024,aubinlequereLLMsResearchTools2024}, to detect AI-related rules across the larger datasets with more precision. 
The approach is based on comparing LLM-produced labels with those produced by human experts and iterating on the LLM prompt until desirable inter-rater reliability (IRR) is achieved between the human and LLM labels.

In order to assess LLM performance on a sample with both positive and negative examples, the first two authors randomly selected $100$ subreddits from the \textbf{Rules Subreddit Set} that matched the AI-keyword regex.
The authors then independently applied a ``Yes/No'' label to each subreddit in response to the following question: \emph{``Do any of this subreddit’s rules explicitly govern the use of AI tools or the content that such tools produce?''}
The two authors achieved strong agreement in their codes $(IRR >.8)$ and only one human coder was used in future iterations.

We used these manually labeled rules as ground truth against which to evaluate the performance of an LLM coder, which we planned to use to label the rest of the data.
We crafted a prompt that combined the human coder instructions with Chain-of-Thought prompting techniques~\cite{wei_chain_thought_2023}, including the recommended eight few-shot, in-context examples, which we selected from the manually labeled sample. 
Using OpenAI's API, we produced LLM labels by prompting \emph{gpt-4o-mini-2024-07-18}, OpenAI's flagship small model at the time, with this prompt and the same input data that were given to the human coders (a subreddit's rules).
We compared the LLM-produced labels with the human-produced ones and calculated an IRR of $.7$, which we deemed insufficient.
Two human coders examined the discrepancies, discussed what they thought the correct label should be, and updated the prompt with a clarifying example.
For example, one rule labeled \textit{NO} by humans but \textit{YES} by the LLM was about bots: \textit{``No bots(No bots): No bots. All bots are banned unless the bot provides exceptional value to /r/ MealPrepSunday.''}
To better align the LLM label with our preferred definition of an AI rule, we added this rule as a negative example in the prompt, with the explanation:``This rule bans bots, not AI. Bots are not necessarily AI.''

To improve IRR we repeated this process of randomly sampling, manually coding, generating LLM labels, comparing the results, and updating the prompt with examples that the model got wrong. 
We performed three iterations of this step, at which point the LLM labels had an IRR of $.95$ with the human labels.
We then generated labels for the entirety of both the \textbf{Longitudinal Subreddit Set} and the \textbf{Rules Subreddit Set} by prompting the \textit{gpt-4o-mini-2024-07-18} model via OpenAI's API.
The final prompt used can be found in Appendix~\ref{app:prompt-detect}.
This process identified the \textbf{AI Rules Subreddit Set} ($N=4,251$) of subreddits with rules governing AI use.

\subsubsection{Identifying an AI Rule Taxonomy}
\label{emergent-rule-types}
In order to better understand the types of rules that subreddits use to govern AI, we identified these rules in our sample and used an iterative inductive coding process to develop a descriptive \textit{AI Rule Taxonomy}.
To identify these rules, we slightly modified the prompt developed and validated in Section~\ref{detection} so that it would flag \textit{individual} rules rather than a set of rules as being AI-related.
Since the prompt required only a slight modification from the prior prompt, and since human coders would be manually reviewing the labels in the next step, we decided that we did not need to validate model performance before tagging the entire sample.
We used this prompt to instruct the \emph{gpt-4o-mini-2024-07-18} model to produce labels for each of the $N=33,674$ individual rules in the \textbf{AI Rules Subreddit Set}, which yielded the \textbf{AI Rules Dataset} ($N=4,458$) of individual rules about AI. 

Next, two of the authors collaborated to identify emergent labels for the rules.
The authors randomly selected $N=25$ rules from this dataset and independently performed affinity mapping~\cite{harboeRealWorldAffinityDiagramming2015} by grouping rules that appeared similar into emergent categories.
The authors then compared results and consolidated their categories into labels that best captured the elements of interest.
Each of the two authors then independently applied this set of labels to another random sample of $N=25$ rules, once again grouping similar rules, including into new emergent groups where the existing labels did not suffice. 
The two authors repeated this process of randomly sampling, independently labeling, collaboratively comparing their results, and consolidating their labels, until no new labels emerged and they were confident that their taxonomy covered the relevant dimensions of variation in the dataset.
The final \textit{AI Rules Taxonomy} contained $28$ labels grouped into four categories: \emph{Rationale}, \emph{Stance}, \emph{Medium} and \emph{Framing}, which are listed in Table~\ref{tab:ruletype-labels} alongside example labels for each category (see Table~\ref{tab:rule_type_emergent} for the full taxonomy and examples rules for each label).
We labeled the entire \textbf{AI Rules Dataset} according to this taxonomy, as we describe next.

\begin{table*}
\centering
\caption{Labels applied to subreddits in the \textbf{Rules Subreddit Set}}
\small 
\begin{tabular}{>{\raggedright}m{2.6cm}>{\raggedright}m{1.9cm}>{\raggedright\arraybackslash}m{2.7cm}>{\raggedright\arraybackslash}m{6cm}}
\hline
\textbf{Label Category} & \textbf{Source} & \textbf{Mutually Exclusive?} & \textbf{Example Labels} \\ \hline
Topic & Fiesler et al. & Yes & Art, Celebrity, Video Games \\ 
Community Archetype & Prinster et al. & No & Content Generation, Topical Q\&A, Social Support \\ Has AI Rule? & New & Yes & Yes, No \\ \hline
\end{tabular}
\label{tab:subtype-labels}
\end{table*}

\subsubsection{Applying Rule Classification Labels}
\label{rule-class-labels}
We used LLMs to label all rules in the \textbf{AI Rules Dataset} with labels from the \textit{AI Rules Taxonomy} and labels from an additional \textit{Rule Type} category provided by Fiesler et al.~\cite{fiesler2018reddit}.
The two first authors collaboratively developed a codebook with instructions for applying these labels.
Fiesler et al. does not provide codebook instructions, so the authors interpreted the examples presented in the study and came up with instructions for how to apply each of the $24$ possible labels.
\textit{Rule Type} labels are not mutually exclusive, so we represented them as a series of binary labels, one for each \textit{Rule Type} label.

The two authors then revisited a sample that they had used for emergent coding and used the instructions to apply \emph{Rule Type} labels.
Next, the authors used their codebook to prompt OpenAI's \emph{gpt-4o-2024-08-06} model to generate the rule-level labels found in Table~\ref{tab:ruletype-labels} for this same sample.
The two authors considered the difference between LLM labels and their own and used the discrepancies to identify parts of their codebook that needed clarification.
The two authors refined the unclear instructions in the codebook and repeated this process iteratively until the LLM labels matched their own with an $IRR > .8$. 
We then prompted the \emph{gpt-4o-2024-08-06} model to generate labels for the entire \textbf{AI Rules Dataset}.
The final prompt used can be found in Appendix ~\ref{app:prompt-rulelabels}.

\subsubsection{Applying Subreddit Classification Labels}\label{sub-labels}
We also sought to label subreddits in the \textbf{Rules Subreddit Set} with details about their community type.
To this end, we devised a process to apply labels from two label categories used in prior HCI literature: Topic labels from Fiesler et al.~\cite{fiesler2018reddit} and Community Archetype labels from Prinster et al.~\cite{prinster2024community}.
In applying labels, we considered a subreddit's \verb|name|, \verb|description|, and \verb|rules| metadata fields.
The label categories for subreddits can be found in Table~\ref{tab:subtype-labels} alongside example labels for each category.

In the first step of the process the first author created a codebook with instructions for labeling (1) the Topic of the subreddit, using as potential labels the $28$ topics provided by Fiesler et al., and (2) which of the five Community Archetypes defined by Prinster et al. apply to the community.
While these prior studies provided useful taxonomies for classification, neither provided a formal codebook with instructions, so the first author developed a codebook with detailed coding instructions based on the examples found in these studies.
Two of the authors used this codebook to independently classify the same sample of $25$ subreddits randomly selected from the \textbf{Rules Subreddit Set}.
The two authors compared their results and calculated IRR, achieving an initial IRR of $.56$, which was deemed insufficient.
To improve IRR, the two authors discussed discrepancies in their codes and updated the codebook instructions where clarity was needed to align their labels.
The two authors repeated this process of randomly sampling $25$ subreddits from the \textbf{Rules Subreddit Set},  independently coding them with the updated codebook, calculating IRR between the two coders, discussing discrepancies between their codes, and iterating on the codebook, until they achieved an IRR of $.88$.
We then used the OpenAI API to generate labels for the most recently coded sample by passing the codebook and subreddit metadata as a prompt to the \emph{gpt-4o-mini-2024-07-18} model.
We compared these LLM-produced labels with the manual labels and found there to be excellent agreement ($IRR=.84$).
We repeated this process on another randomly sampled $25$ subreddits.
On this new sample the manual labels again agreed with the LLM labels with excellent agreement ($IRR=.88$).
We then used the OpenAI API to prompt the \emph{gpt-4o-mini-2024-07-18} model to generate labels for each subreddit in the \textbf{Rules Subreddit Set}.
The final prompt used can be found in Appendix ~\ref{app:prompt-sublabels}.

\section{Findings}
We group our results by research question and present in sequence our findings on: (1) the changes in the prevalence of community rules governing AI use, (2) the types of communities that have these rules, and (3) details about the rules themselves.

\begin{table*}[t]
\caption{Prevalence of rules and rules governing the use of AI in the \textbf{Longitudinal Subreddit Set} for (1) all subreddits and (2) the top 1\% largest subreddits, ranked by subscriber count at the time of the 2024 crawl.}
\begin{tabular}{lllll|l}
\multicolumn{1}{l}{\multirow{2}{*}{}}                            & \multicolumn{2}{l}{\textbf{July 2023}}                             & \multicolumn{2}{l}{\textbf{November 2024}}                                                  \\
\multicolumn{1}{l}{}                                             & \multicolumn{1}{l}{N}      & \multicolumn{1}{l}{\% of sample}     & \multicolumn{1}{l}{N}      & \multicolumn{1}{l|}{\% of sample}     & \textbf{\% Change} \\ \hline
\textbf{All subreddits in dataset} ($N=227,737$)               &  & &  &          \\
\multicolumn{1}{l}{had rules at time of crawl}                   & \multicolumn{1}{l}{66,505} & \multicolumn{1}{l}{29.2\%} & \multicolumn{1}{l}{71,862} & \multicolumn{1}{l|}{31.6\%} & \textbf{+8.1\%}    \\
\multicolumn{1}{l}{had AI rules at time of crawl}                & \multicolumn{1}{l}{1,348}  & \multicolumn{1}{l}{.6\%}   & \multicolumn{1}{l}{2,808}  & \multicolumn{1}{l|}{1.2\%}  & \textbf{+108.3\%}  \\ \hline
\textbf{Top 1\% largest subreddits} ($N=2,278$) & & &  & & \\
\multicolumn{1}{l}{had rules at time of crawl}                   & \multicolumn{1}{l}{2139}   & \multicolumn{1}{l}{93.9\%} & \multicolumn{1}{l}{2197}   & \multicolumn{1}{l|}{96.4\%} & \textbf{+2.7\%} \\
\multicolumn{1}{l}{had AI rules at time of crawl}                & \multicolumn{1}{l}{192}    & \multicolumn{1}{l}{8.4\%}  & \multicolumn{1}{l}{390}    & \multicolumn{1}{l|}{17.1\%} & \textbf{+103.1\%} \\ \hline
\end{tabular}
\label{tab:ai-rules-presence}
\end{table*}

\subsection{RQ1: How Has the Prevalence of AI Rules Changed Over Time?}\label{rq1}

\begin{figure}[b]
    \centering
    \includegraphics[width=1\linewidth]{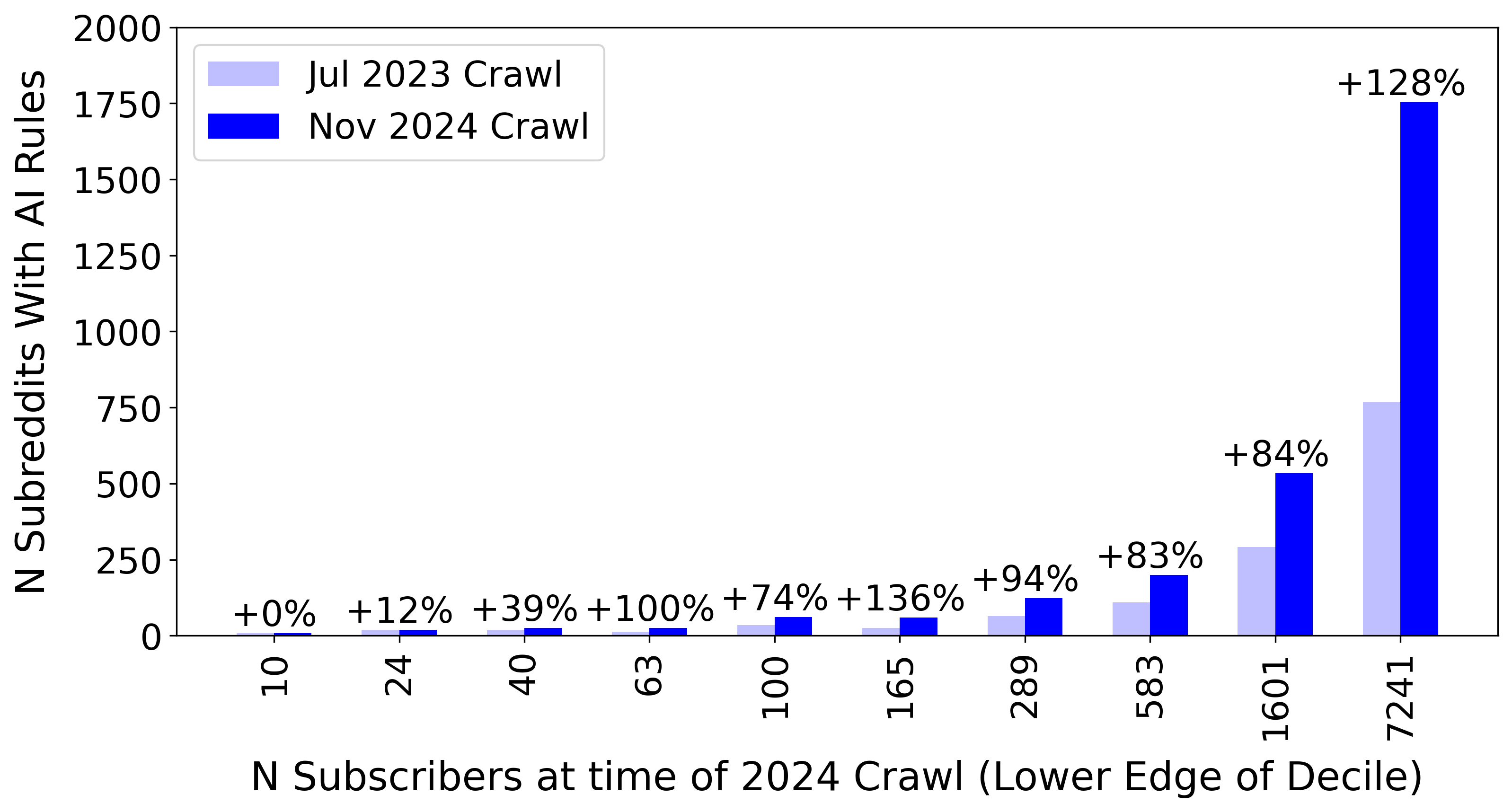}
    \caption{Number of subreddits in the \textbf{Longitudinal Subreddit Set} with AI rules. 
    Subreddits are bucketed into deciles based on their subscriber count at the time of the second crawl. 
    Labels above the bars indicate percent change between crawls.}
    \label{fig:airules_increase2}
    \Description{The image contains a bar chart showing the number of AI-related rules in subreddits during two different crawls, with subreddits bucketed into deciles by subscriber count.
The Y-axis represents the number of subreddits and the X-axis is subscriber count decile at the time of the second crawl (measured by the lower edge of each decile).
Each decile has two bars above it: the left bar represents the 2023 crawl and the right represents the 2024 crawl.
The number of subreddits with AI rules steadily increases as subscriber count grows. The largest decile (7,241) has the highest number, above 1750 during the second crawl, while the smallest decile (10) has a number near 0.}
\end{figure}

Table~\ref{tab:ai-rules-presence} shows the results of the detection task described in Section~\ref{detection} on our \textbf{Longitudinal Subreddit Set}.
We find that during the 2023 crawl, $N=1,348$ subreddits had rules governing the use of AI ($.6\%$).
By the time of our 2024 crawl, this number increased by $108.3\%$ to $N=2,808$ ($1.2\%$).
Rules in general, and specifically rules governing the use of AI, are more common in larger subreddits.
For example, $8.4\%$ of the largest $1\%$ of subreddits by subscriber count had AI rules during our first crawl.
This percentage increased to $17.1\%$ in our second crawl.
Figure \ref{fig:airules_increase2} provides a more detailed demonstration of the differences in the prevalence of AI rules and the rates of increase in such rules among subreddits with different numbers of subscribers.
The figure shows the number of subreddits in the \textbf{Longitudinal Subreddit Set} with AI rules at the time of each crawl.
In order to expose the variation across subreddits of different sizes, we bucket subreddits along the X-axis into deciles based on their number of subscribers at the time of the second crawl (the labels represent the lower edge of the decile).
Each decile contains approximately 22,773 subreddits.
For each decile we plot two bars representing the number of subreddits with AI rules during the first crawl (left bar, lighter blue) and the second crawl (right, darker blue).
The number above the bars represents the percent change between these two crawls.
For example, the rightmost pair of bars shows that in the largest decile of subreddits, about $750$ had AI rules during the first crawl, and that this number increased by $128\%$ to about $1,750$ at the second crawl.
Remember that each decile contains the same total number of subreddits. 
The figure thus shows that subreddits with more subscribers are more likely to have AI rules, and that communities of all sizes saw increases in the prevalence of AI rules between our two crawls, though that increase was more prominent for larger communities.

For comparison, we additionally analyzed the prevalence of AI rules in subreddits that are part of the more recent \textbf{Broad Subreddit Set} and that were created \textit{after} our first crawl and were thus excluded from our \textbf{Longitudinal Subreddit Set}.
We identified $N=30,962$ English language subreddits in the \textbf{Broad Subreddit Set} that were created after the first crawl but were at least a month old during the second crawl. 
Of these newly created subreddits, $N=467$ had AI rules ($1.5\%$),
which is higher than the percentage of subreddits with AI rules during the second crawl in the \textbf{Longitudinal Subreddit Set} ($1.2\%$).
While this difference may be due to the fact that new subreddits have different characteristics, our separate analysis in the following section shows that new subreddits are heavily featured in the distribution of subreddits with AI rules in our data.

\subsection{RQ2: What Types of Communities Have Rules Governing AI Use?}
\label{rq2}
In this section, we examine the characteristics of communities that have rules governing the use of AI. 
To analyze these characteristics, we ran two sets of analysis, one on the \textbf{Broad Subreddit Set} and another on the \textbf{Rules Subreddit Set}.
The results from our \textbf{Broad Subreddit Set} analysis show that larger subreddits are most likely to have explicit rules regulating AI, compared to subreddits with no rules, as well as those with rules but without AI rules. 
The results from our \textbf{Rules Subreddit Set} analysis show that among subreddits with any rules, subreddits with AI rules are more likely to be associated with the \textit{Art} or \textit{Celebrity} topics than subreddits without AI rules.

\subsubsection{\textbf{Broad Subreddit Set} Analysis}\label{rq2-broad_subreddit_set}
To determine which characteristics of subreddits specifically predict the presence of AI rules, as opposed to rules in general, we analyzed three groups of subreddits in the \textbf{Broad Subreddit Set}: subreddits with no rules, subreddits with AI rules, and subreddits with rules but no AI rule.
Across these three groups, we looked at subreddits' \textit{subscriber count} and \textit{age} (in years).
We apply a Kruskal-Wallis test and find evidence that the distributions of both characteristics are different across the three groups (significant with $p<0.01$).
Of the three groups, subreddits with AI rules have the most subscribers: these subreddits had a median subscriber count of 5,783 (distribution on $\log_{10}$ scale with x-axis $[1,8]$=
\begin{sparkline}{10}
 \sparkspike 0.0 0.423
 \sparkspike 0.1 0.582
 \sparkspike 0.2 0.764
 \sparkspike 0.3 0.867
 \sparkspike 0.4 1.0
 \sparkspike 0.5 0.872
 \sparkspike 0.6 0.427
 \sparkspike 0.7 0.211
 \sparkspike 0.8 0.042
 \sparkspike 0.9 0.027
\end{sparkline}
), compared to a median count of 354 for subreddits with rules but no AI rules (distribution=
\begin{sparkline}{10}
 \sparkspike 0.0 0.915
 \sparkspike 0.1 1.0
 \sparkspike 0.2 0.836
 \sparkspike 0.3 0.639
 \sparkspike 0.4 0.5
 \sparkspike 0.5 0.29
 \sparkspike 0.6 0.115
 \sparkspike 0.7 0.036
 \sparkspike 0.8 0.012
 \sparkspike 0.9 0.002
\end{sparkline}
), and a median of 90 for subreddits without any rules (distribution=
\begin{sparkline}{10}
 \sparkspike 0.0 0.889
 \sparkspike 0.1 1.0
 \sparkspike 0.2 0.644
 \sparkspike 0.3 0.297
 \sparkspike 0.4 0.135
 \sparkspike 0.5 0.047
 \sparkspike 0.6 0.01
 \sparkspike 0.7 0.002
 \sparkspike 0.8 0.0
 \sparkspike 0.9 0.0
\end{sparkline}
). 
Subreddits with AI rules are also older: the median age of these subreddits was $4.64$ years (distribution with x-axis $[0, 18]$=
\begin{sparkline}{10}
 \sparkspike 0.0 1.0
 \sparkspike 0.1 0.385
 \sparkspike 0.2 0.302
 \sparkspike 0.3 0.175
 \sparkspike 0.4 0.159
 \sparkspike 0.5 0.226
 \sparkspike 0.6 0.445
 \sparkspike 0.7 0.272
 \sparkspike 0.8 0.229
 \sparkspike 0.9 0.001
\end{sparkline}), compared to a median age of $3.97$ years for subreddits with rules but no AI rules (distribution=
\begin{sparkline}{10}
 \sparkspike 0.0 1.0
 \sparkspike 0.1 0.817
 \sparkspike 0.2 0.857
 \sparkspike 0.3 0.371
 \sparkspike 0.4 0.264
 \sparkspike 0.5 0.211
 \sparkspike 0.6 0.253
 \sparkspike 0.7 0.163
 \sparkspike 0.8 0.105
 \sparkspike 0.9 0.0
\end{sparkline}), and a median age of $4.09$ years for subreddits without any rules (distribution=
\begin{sparkline}{10}
 \sparkspike 0.0 1.0
 \sparkspike 0.1 0.953
 \sparkspike 0.2 0.735
 \sparkspike 0.3 0.382
 \sparkspike 0.4 0.379
 \sparkspike 0.5 0.446
 \sparkspike 0.6 0.366
 \sparkspike 0.7 0.121
 \sparkspike 0.8 0.041
 \sparkspike 0.9 0.0
\end{sparkline}).
However, we note that while the median age of subreddits is largest in the group with AI rules, that group also had the largest percentage of its subreddits in the smallest decile of the age distribution, compared to the two other groups.
This is in line with the findings discussed at the end of Section~\ref{rq1}, that the newest subreddits are more likely to have AI rules.
We interpret the relationship between age and AI rules further in the next paragraph.

To validate our findings about differences in the subscriber count and age distributions between subreddit groups, we next ran a multinomial logistic regression on this same set of subreddits.
We used as input variables \textit{subscriber count}\footnote{We log transform \textit{subscriber count} to account for the variable's skewed distribution.} and \textit{age} (in years), in an attempt to control for any interaction between the two.
The dependent variable took one of three mutually exclusive values based on the rules associated with the subreddit during our crawl: \textit{has no rules}, \textit{has rules but no AI rule}, and \textit{has AI rule}.
This model found significant correlations between both input variables and the type of rules present.
Compared to subreddits with rules but no AI rule, an increase in subscriber count increased the odds of having AI rules ($\beta=0.32, p<0.01$) while an increase in age decreased the odds ($\beta=-0.035, p<0.01$). 
Still using as baseline the subreddits with rules but no AI rule, the model output shows that an increase in subscriber count decreased the odds of having no rules ($\beta=-0.41, p<0.01$) while an increase in age increased these odds ($\beta=0.11, p<0.01$). 
These results further support the finding from our distribution comparison that larger subreddits are more likely to have AI rules.
These results also provide a more nuanced understanding of the relationship between subreddit age and the presence of AI rules.
While the distribution comparison found that subreddits with AI rules have the largest median age, the regression results suggest that this is partially due to an interaction between subscriber count and age.
When controlling for subscriber count, the regression results suggest that subreddits with AI rules are actually more likely to be \textit{newer} than both those with no rules and those with rules but no AI rules.

\subsubsection{\textbf{Rules Subreddit Set} Analysis}\label{rq2-rules_subreddit_set}
\begin{table*}
\small
\caption{
Differences in the prevalence of labels for \textbf{Rules Subreddit Set} subreddits ($N=99,969$) with and without AI rules.
Out of the 28 topic labels,  we show the ten with the largest absolute differences.
}
\begin{tabular}{llllll}
& Label & Subreddits w/out AI Rule & Subreddits w/ AI Rule & Difference    \\
\hline
\multirow{5}{*}{Community Archetype} &Content generation   & 51.68\% & 73.75\% & 22.07*** \\
& Affiliation with an entity  & 57.47\% & 63.3\% & 5.83*** \\
& Topical question and answer &  7.29\% & 6.52\% & -0.78 \\
& Learning and perspective broadening  & 20.52\% & 16.61\% & -3.91*** \\
& Social support  & 10.29\% & 4.38\% & -5.91*** \\
\hline
 \multirow{10}{*}{Topic} &
Celebrity  & 3.87\% & 18.33\% & 14.45*** \\
& Art  & 2.36\% & 12.99\% & 10.63*** \\
& Entertainment  & 19.01\% & 22.96\% & 3.95*** \\
& Politics  & 2.21\% & 0.54\% & -1.67*** \\
& Business and Organizations  & 2.71\% & 0.99\% & -1.73*** \\
& Games  & 6.97\% & 4.68\% & -2.29*** \\
& Local  & 4.22\% & 1.11\% & -3.11*** \\
& Hobby   & 8.07\% & 4.7\% & -3.36*** \\
& Humor   & 9.75\% & 3.32\% & -6.43*** \\
& Support   & 8.98\% & 2.52\% & -6.46*** \\
\hline
\multicolumn{5}{l}{Note: $^*p<0.1, ^{**}p<0.05, ^{***}p<0.01$}
\end{tabular}
\label{tab:balance-test}
\end{table*}

To determine which characteristics of subreddits with rules predict the presence of AI rules, we analyze two groups of subreddits in the \textbf{Rules Subreddit Set}: subreddits with rules about the use of AI, and those without such rules. 
Across these two groups, we consider the presence of the various topic and Community Archetype labels assigned to the subreddits, and implement Bonferroni correction to adjust for multiple comparisons. 
We first perform difference-in-proportions tests, using the Chi-squared test for significance.
Table~\ref{tab:balance-test} presents the results of this test.
Notice that the topic labels are mutually exclusive, but Community Archetype labels are not. 
We find that there is a large and significant difference between subreddits with the \textit{Content Generation} label; subreddits with AI rules are about $22$ percentage points more likely to have this label. 
In contrast, subreddits with AI rules are about $6$ percentage points \textit{less likely} to have the \textit{Social Support} Community Archetype label.
There are also significant differences among the topic labels: out of all the mutually exclusive labels, \textit{Art} and \textit{Celebrity} are more common in subreddits with AI rules, encompassing about a third of all subreddits with AI rules compared to only $6\%$ of subreddits with no AI rules.

To validate these results we additionally run a logistic regression model.
We use all subreddit topic and Community Archetype labels, a subset of which are shown in Table~\ref{tab:balance-test}, as input variables and use the presence of AI rules as the dependent variable.
We get results in line with the findings from our difference-in-proportions tests.
For example, the topic labels most predictive of AI rules are \emph{Art} and \emph{Celebrity} ($\beta = 1.4, p<0.01$ and $\beta = 1.69, p<0.01$), supporting the analysis above.
In terms of Community Archetype labels, the label that is most predictive of AI rules is \emph{Content Generation} ($\beta=0.96, p<0.01$), which also supports the above analysis.

Some illustrative examples provide additional context for interpreting these statistical observations.
For example, consider a Wes Anderson subreddit whose purpose is to discuss ``photos that accidentally share Wes Anderson's unique style." 
Our labeling procedure generated both the \textit{Art} topic and the \textit{Content Generation} Community Archetype for this subreddit. 
The language of this subreddit's AI rules reflects a distinct concern with the veracity of posted \textit{images}, rather than concerns about authentic interactions.
The subreddit's rules state that ``Your photos don't have to be images taken by you, but they must be real photos.'' 
Similarly, subreddits labeled with the \textit{Celebrity} topic tend to group AI rules with bans of photoshop, white borders, or low-resolution images, all of which are concerned with the format and quality of the subreddit's content. 
In contrast, subreddits that we labeled with the \textit{Social Support} community archetype like r/introverts -- a subreddit that does \textit{not} have an AI rule -- emphasize in their rules that they care about ``authentic connections.'' 
We further explore the implications of these observations in Section~\ref{discussion}.

\subsection{RQ3: What Types of Rules Exist for Governing AI?}
While our previous analyses explored what types of communities are likely to explicitly regulate AI, we also consider differences in \textit{how} communities regulate AI. 
We examine the frequencies of the labels described in Sections~\ref{emergent-rule-types} and \ref{rule-class-labels} for rules in the \textbf{AI Rules Set} ($N=4,458$).
Table~\ref{tab:rule_type_summary} shows the frequency of rules labeled with~\citet{fiesler2018reddit}'s Rule Type labels, while Table~\ref{tab:rule_type_emergent} presents the frequency of rules labeled with the \textit{AI Rules Taxonomy} produced in Section~\ref{emergent-rule-types}. 
The tables show the percentage of rules in the set that were tagged with each label.

Comparing the results in Table~\ref{tab:rule_type_summary}  to a similar analyses of the prevalence of the same labels in a \textit{general} subreddit rules dataset---performed by Fiesler et al.~\cite{fiesler2018reddit}---illuminates some interesting differences between AI rules and general subreddit rules.
Fiesler et al.'s analysis of general subreddit rules found that \textit{Prescriptive} rules, or rules that tell users what to do, occurred in about equal numbers as \textit{Restrictive} rules that tell users what \textit{not} to do\footnote{These labels are not mutually exclusive: a rule could be tagged as both \textit{Prescriptive} and \textit{Restrictive} if it contains elements of both style.}.
However, in our \textbf{AI Rules Set}, $96.84\%$ of rules were tagged as \textit{Restrictive} while only $67.7\%$ were tagged as \textit{Prescriptive}.
A \textit{Restrictive} rule might include statements like ``No AI (chatGPT etc.)'' or ``No AI generated content'' while a \textit{Prescriptive rule} might instead tell users whether to cite if AI is used. 
This pattern further bears out in the data for the \textit{AI Rules Taxonomy}: the restrictive \textit{Unqualified Bans} (e.g., ``No AI-generated content'') is the most common label ($55.23\%$ of AI rules) in the \textit{AI Rule: Stance} label category in Table~\ref{tab:rule_type_emergent}.
It seems that many subreddits may find it easier to simply ban the use of AI outright, rather than trying to develop policies that can accommodate it under certain circumstances (e.g., requiring the disclosure of AI use). 
Still, we do see evidence of some communities adopting stances other than outright bans, like the $24.23\%$ of AI rules labeled \textit{Stance: Qualified Ban} and the $17.95\%$ labeled \textit{Stance: Disclosure}. 

Overall, we observe that a high number of AI rules are related to images. 
Table~\ref{tab:rule_type_summary} shows that $46.55\%$ of rules were tagged with the \textit{Images} label from \citet{fiesler2018reddit}, which according to the codebook we developed indicates that the rule ``is about posting visual content like images, photos, or paintings''.
For comparison, \citet{fiesler2018reddit} tagged less than $8\%$ of the rules in their dataset with this label.
We see a similar trend in the labels from our \textit{AI Rules Taxonomy}: Table~\ref{tab:rule_type_emergent} indicates that $31.4\%$ of AI rules are labeled as \textit{Medium: Images} and $30.26\%$ are labeled \textit{Medium: Art}. The union of these two sets accounted for $57.02\%$ of AI rules in the data. 
This observation further supports the finding in Section~\ref{rq2} that subreddits with \textit{Art} and \textit{Celebrity} topics are more likely to have AI rules, as these are usually image-sharing subreddits. 

Tables~\ref{tab:rule_type_summary} and ~\ref{tab:rule_type_emergent} also demonstrate that many AI rules on Reddit focus on concerns about content quality. 
In Table~\ref{tab:rule_type_summary}, around $28\%$ of rules are labeled as relating to \textit{Low quality content}; in Table~\ref{tab:rule_type_emergent}, $13\%$ and $11\%$ respectively are labeled as \textit{Rationale: Low quality} and \textit{Rationale: Low effort}, meaning that these rules explicitly stated these concerns as rationales for regulating AI use. 
Rules about \textit{Low quality content} emphasize the importance of ``creative merit'' and content that takes more than ``only a few minutes of effort.''
These rationales are more prevalent than rationales based on spam, ethics, or accuracy. 
The focus on quality may suggest that explicit AI regulation is driven by practical concerns about content and reputation, not ideological objections over the use of AI more broadly. 
This observation echoes the ``content quality concerns'' shared by moderators in our earlier qualitative work\cite{lloyd_there_2023}.

\begin{table}
\small
\caption{Percentage of rules in the \textbf{AI Rules Set} ($N=4,458$) labeled with Fiesler et al.'s non-mututally exclusive Rule Type labels.}
\begin{tabular}{ll}
                                    \textbf{Rule Type}  & \textbf{\%}  \\\hline
Content and behavior & 99.73 \\
Restrictive & 96.84 \\
Prescriptive & 67.7 \\
Images & 46.55 \\
Consequences, moderation, and enforcement & 32.48 \\
Low quality content & 27.77 \\
Post format & 19.25 \\
Copyright and piracy & 16.49 \\
Links and outside content & 10.54 \\
Spam & 9.02 \\
Off topic & 7.11 \\
Reposting & 5.88 \\
NSFW & 5.59 \\
Advertising and commercialization & 3.93 \\
Reddiquette and sitewide rules  & 1.05 \\
Politics & 1.03 \\
Harassment & 0.96 \\
Spoilers & 0.7 \\
Hate speech & 0.61 \\
Doxxing and personal information & 0.54 \\
Trolling & 0.43 \\
Voting & 0.18 \\
Personal army & 0.13 \\
Personality & 0.04 \\
\hline
\end{tabular}
\label{tab:rule_type_summary}
\end{table}

\begin{table*}
\renewcommand{\arraystretch}{1.3}
\fontsize{7.5}{8.5}\selectfont

\caption{Percentage of rules in the \textbf{AI Rules Set} ($N=4,458$) with non-mutually exclusive labels from our \textit{AI Rules Taxonomy}. Examples come from the data but have been shortened for brevity. See the tagging prompt in Appendix \ref{app:prompt-rulelabels} for label definitions.}
\begin{tabular}{p{1.9cm}p{1.75cm}p{1.25cm}>{\itshape}p{8.75cm}} \\ \textbf{Rule Category}
& \textbf{Rule Label}  & {\normalfont\bfseries \% Of Rules } & {\normalfont\bfseries Example}\break \\\hline
\multirow{18}{*}{\textbf{Rationale}} & Low quality                  & 13.19 & Low quality content will be removed. This includes AI generated images and other AI media.\\
& Inauthentic                & 11.93 & NO FAKES OR OVERLY SHOPPED PHOTOS - may result in a ban, serious offense. Images must be posted as published except for Quality AI-enhanced images. We ask that those titles include "Enhanced."\\

& Low effort                 & 11.06 & AI generated images or Chatgpt text content will be removed as "low effort" content. Please do not post AI generated content on this subreddit.\\

& Original content           & 7.0  & AI generated text does not qualify as an original source.\\
& Off topic                  & 3.1  & Content generated by a computer algorithm is off topic and unwanted.\\
& Spam                      & 3.25  & Any form of spam / blogspam is not allowed.  This includes links to book reviews, AI generated content,  bots, social media channels and any affiliate links.\\
& Ethics                     & 2.98  & Given Miyazaki's vehement opposition to AI and the ethical issues with AI generated art, posts that feature content generated by an AI model will be removed.\\
& Policy                     & 1.37  & Deep fakes violate reddit TOS. You will be banned if you post one.\\
& Misleading                 & 1.7  & AI altered videos are misleading and will be removed.\\
& Inaccurate                 & 0.83  & AI is awesome, but when used to answer technical questions it often gives incorrect or sometimes dangerous recommendations.\\
& Inhuman                    & 0.9  & We believe Hip hop is human made, and for that reason we've decided to ban AI generated content.\\
\hline
\multirow{6.5}{*}{\textbf{Stance}} & Unqualified ban             & 55.23 & AI-generated content of any kind is forbidden.\\
& Qualified ban                & 24.23 & AI Generated Fan Art is also not allowed, UNLESS it has been substantially edited by a human.\\
& Disclosure                   & 17.95 & If posting AI-generated art, specify in the title that it is AI-created, include what service used.\\
& Enabling                     & 18.51 & Recipes written by AI are ok but you must: 1) make the salsa, and 2) post a real photo of the completed salsa.\\
& Requiring                    & 1.1 & Only post images generated by Ai, that's it\\
\hline
\multirow{13}{*}{\textbf{Medium}} & Content  & 52.74 & On r/NFT, content that is fully generated by AI and not modified in any way shape or form is prohibited and will result in a ban from the community.\\
 & Images                        & 31.4 & Please no DALL·E mini or other AI generated images.\\
 & Art                           & 30.26 & No AI art allowed, as per our content policy.\\
 & Fakes                        & 15.79 & No fake images of Leah Willamson are permitted, either photoshopped or AI generated\\
 & Unspecified                         & 5.36 & No AI - https://www.gofundme.com/f/protecting-artists-from-ai-technologies\\
 & Text                               & 6.59 & This includes both AI generated images/artwork and text generated by tools such as ChatGPT.\\
 & Deepfake                      & 4.6 & Deepfakes or other types of fakes are not allowed here, you will be banned for posting them.\\
 & Bots                             & 3.34 & Karma-farming with ChatGPT bots, Malware, phishing, spam or scam will result in your account being banned without warning.\\
 & Videos                         & 3.95 & No low effort AI content or Videos.\\
\hline
\multirow{5}{*}{\textbf{Framing}} & Generate                       & 53.88 & AI-generated content of any kind is forbidden.\\
& Assist                         & 16.08 & Traditional paintings and drawings only. No videos, screenshots, photography, audio, multi-panel comics, content generated or AI-assisted submissions.\\
& Unspecified                        & 12.02  & The removal of low quality and, AI submissions, are subject to remove at mod discretion.
\end{tabular}
\label{tab:rule_type_emergent}
\end{table*}

Since the \textit{Art} and \textit{Celebrity} topics featured prominently in Section~\ref{rq2}, we also separately examined patterns among Rule Types within each of these Topics. 
Among subreddits with the \textit{Art} Topic label, we find a higher share of AI rules with the \textit{Copyright and Piracy} label ($25.54\%$ vs $16.49\%$ baseline).
These \textit{Copyright and Piracy} rules primarily focus on proper attribution and posting original content.
Among \textit{Celebrity} subreddits, we see that the \textit{Low Quality Content} and \textit{Format} Rule Type labels are much more common ($58.32\%$ and $55.32\%$ vs baselines of $27.77\%$ and $19.25\%$).
The \textit{Format} label means that references to AI often co-occur with regulations about the format of posts --- such as statements about how posts should be titled or specifications about minimum word counts.
In terms of the explicit rationales given for the AI rules, almost half of AI rules in \textit{Celebrity} subreddits were labeled \textit{Rationale: Inauthentic} and $34.79 \%$ were labeled \textit{Rationale:~Low Quality}, compared with $11.93\%$ and $13.19\%$ respectively in the set of all AI rules. 
Additionally, $77\%$ of AI rules from \textit{Celebrity} subreddits explicitly govern the use of ``Fakes'' (labeled with \textit{Medium: Fakes}), compared to $15.79\%$ in the wider set. 
Upon closer examination, rules about ``Fakes'' in \textit{Celebrity} subreddits appear to be referring to heavily-edited photos (these rules separately make references to ``Deepfakes'', which we distinguish with a different label).

\section{Discussion}\label{discussion}
Our quantitative study offers three main research contributions: (1) a large-scale analysis of rules about AI use in Reddit communities, (2) a taxonomy for classifying these rules, and (3) a dataset of subreddits, their metadata, and their published community rules at two distinct points in time.
Our study aims to inform and support community self-moderation and draws inspiration from Seering's second ``guiding question'' for the HCI research community: \emph{``What are the processes of context-sensitivity in online community moderation, and how might they be better supported?''}~\cite{seeringSelf}.
The sudden arrival of generative AI poses unique challenges for online communities~\cite{lloyd_there_2023} that HCI research is uniquely positioned to help address.
Our study attempts to measure at scale the prevalence and variety of rules that online communities have enacted in response to this new technology.
We hope that our data and analysis can add additional context to qualitative studies into online communities' responses to AIGC, as well as to general knowledge about online community self-moderation practices.
Below we put our findings in dialogue with prior work and discuss design implications for social platforms.

\subsection{The Role of Community Rules}\label{role}
While we focus our analysis on rules about AI use, one of our primary contributions is a dataset of community rules that the HCI research community can use to answer a wider range of questions about community self-moderation.
As far as we know, this is the first dataset of its kind, as prior studies of self-governing online communities' rules~\cite{fiesler2018reddit,Nicholson23,Cai21,hwangRulesRuleMakingFive2022,Reddy23} have not made their rules datasets public.
Large-scale, quantitative analysis of these rules offers a view into platform-wide patterns that complement qualitative studies of online community self-moderation.
Our AIGC-focused analysis finds that subreddit communities, especially those with more subscribers, are changing their rules to respond to generative AI.
But what do these new AI rules tell us about the experience of participating in or moderating these communities?

Past research into community self-moderation has found that community rule changes are most often enacted, ``in response to unexpected incidents'' that impact a community~\cite{seeringModeratorEngagementCommunity2019}.
Our earlier, qualitative work on the experience of moderating AIGC~\cite{lloyd_there_2023} provides evidence that AI rules are created for the same reason, as many moderators report problematic AI use in their communities as a precursor to their formal rules. 
This mechanism of rule creation suggests that a community's rules are not a perfect proxy for its values, as community values do not need to be codified into rules unless there is some incident that surfaces community misalignment.
Seen in this light, the communities with AI rules in our dataset are not necessarily the communities with the strongest stances on AI, but those where some misalignment about AI use emerged between community members. 
Viewing rules as evidence of past misalignment could explain why \textit{Social Support} subreddits were negatively associated with AI rules in our analysis: perhaps the members who participate in these communities are more aligned in their views towards AI, so that explicit rules are not necessary.
In general, we suspect that larger, less personal communities, with looser ties between members, may be more likely to rely on formal rules to maintain content quality and standards, as opposed to smaller, closer-knit communities which can use more informal methods, such as community norms.

It is worth noting that while rules may indicate some past community misalignment, they do not necessarily tell us how rules shape behavior in practice.
For example, prior work has shown that humans cannot reliably detect AIGC~\cite{jakeschHumanHeuristicsAIgenerated2023}, which raises the question of how enforceable AIGC rules are. 
At the same time, it is unclear whether or not AIGC rules even \textit{need} to be enforced: past work on community rules suggests that they can shape behavior by clarifying norms~\cite{matias2019preventing}, even without an effective enforcement mechanism.
Still, our prior work shows that some communities with AI rules \textit{are} attempting to enforce them~\cite{lloyd_there_2023}. 
We see the prevalence of AI rules, especially in the largest subreddits, as evidence that HCI research should explore design solutions that can help with their enforcement.

One way that HCI research can empower communities to effectively enforce their AIGC policies is by offering tools for policy design.
While some online communities are enacting top-down AI policies~\cite{stackoverflow24}, Reddit has so far left this decision to individual communities.
Such an approach empowers community moderators who work and interact closely with their community~\cite{cullenPracticingModerationCommunity2022}, resulting in a quick feedback loop that better positions them to craft context-sensitive policies~\cite{seeringSelf}.
Still, opportunities exist for HCI research to help communities enact effective policies, both by offering empirical evidence of the effects of certain types of rules, as well as new systems for policy development~\cite{Zhang20}.
Future work could use our rules dataset to find similar communities that differ in their policy choices, which could be used as a natural experiment to study the effects of different rules. 
Studying the effects of policies may uncover general knowledge about which kinds of rules (such as ``prescriptive'' vs ``restrictive'') encourage community engagement and which cause members to leave for other communities. 
Platforms could use this knowledge to offer moderators a reasonable set of ``default'' rules at community creation time, based on the type of community being created.
Alternative systems for policy development from past HCI research have explored ways of engaging more members of a community, beyond just moderators, in the policy creation process~\cite{Zhang20}.
We see these more democratic systems of rule development as promising at a time when norms around AI use are still in flux.

\subsection{Supporting Community Self-Governance in the Age of Generative AI}\label{discussion_self_governance}
Comparing our findings to those from an earlier large-scale analysis of Reddit rules conducted by Fiesler et al.~\cite{fiesler2018reddit} provides insight into what is different about AI rules. 
We found that rules governing AI usage are more likely to be about images than the general rules analyzed in Fiesler et al. 
Together with the finding that images are the medium most referenced in AI rules, this finding suggests that the use of AI images may be the most common cause of community misalignment about AI norms. 
Our findings that \textit{Restrictive} AI rules are more common than \textit{Prescriptive} ones seems to suggest that, at least at this point in time, communities appear more interested in clarifying how AI \textit{cannot} be used than how it can.
Still, there is some evidence of flexibility in these stances, and it will be interesting to observe if community stances towards AI grow more or less flexible over time.

These rules---and norms about AI use---will continue to change over time.
Past HCI studies have shown that a strong social identity can help close-knit online communities navigate the challenging task of developing new norms around controversial technology use~\cite{fiesler19}, and we suspect the same will be true with generative AI.
So far our data shows that AI rules have become more prevalent, but these rules may disappear if AI use becomes widespread to the point that it is unremarkable, or where bans have negative impacts on community engagement.
Two of the top three most common rationales in the AI rules in our data were \textit{``Inauthentic''} and \textit{``Low effort''}, which seems to suggest that some communities perceive that the use of AI cheapens the social interactions that occur within them.
These rationales for AI rules echo the findings from prior studies into the use of automated tools in interpersonal communication~\cite{moradiSociotechnicalChangeTracing2024,rae_effects_2024}.
However, a closer look into our data reveals that these rules are more often about the authenticity of \textit{content}, rather than interpersonal interactions. 
We will discuss these concerns about content authenticity more in the next paragraph.
For communities who are concerned with authentic interactions, we suspect that even if communities cannot effectively ban AIGC, those that care about gate-keeping~\cite{keeganEgalitariansGateOnesided2010} will seek out other ways for users to demonstrate their authenticity, similar to those that have developed in some anonymous online communities~\cite{bernstein4chanAnalysisAnonymity2011}. 
Platform designers can help by offering communication tools that encourage authenticity, maybe borrowing from the above, as well as from recent work on \textit{effortful communication}~\cite{Kelly17,Zhang22}.

Our results complement our earlier qualitative investigation into moderators' experience of enforcing AIGC rules~\cite{lloyd_there_2023}.
That prior qualitative, interview-based study identified three main categories of moderator concern about how AIGC will impact their communities: concerns about content quality, social dynamics, and governance processes. 
Of these three concerns, the platform-wide quantitative analysis in this current work shows that concerns about content quality are the most prevalent in subreddit community rules.
Our results in Section~\ref{rq2-rules_subreddit_set}, where we found that AI rules are more common in \textit{Content Generation}, \textit{Art}, and \textit{Celebrity} subreddits, suggest that rules about content quality might be most useful as ways to maintain standards in content-sharing communities. 
These communities tend to be less personal, with looser ties between members. 
We suspect that rules containing social dynamics concerns, highlighted in the earlier work~\cite{lloyd_there_2023}, are less prevalent in our data because of the phenomenon that we discuss in Section~\ref{role}, where desired social dynamics may not need to be codified into rules because the communities that care about them may be naturally better aligned.  

Additionally, our quantitative results provide further evidence of the qualitative finding that moderators' concerns about AIGC are based on their communities' values~\cite{lloyd_there_2023}. 
While rules usually do not explicitly state a community's values, we can often infer them. 
For example, it is not a coincidence that \textit{Copyright and Piracy} AI rules are more common in \textit{Art} subreddits.

Finally, we note one interesting difference between the findings in our current study and those from our earlier qualitative work~\cite{lloyd_there_2023}.
While the majority of concerns raised in moderator interviews pertain primarily to text~\cite{lloyd_there_2023}, rules about images are more common in our data.
Helping communities respond to AI-generated text as opposed to AI-generated images is a different challenge with different potential solutions.
For example, images may be more likely to be robustly watermarked~\cite{dathathriScalableWatermarkingIdentifying2024} or to have provenance metadata~\cite{Feng23}, which platforms could clearly communicate to viewers.

This current work is not without limitations. 
One limitation is that this study is based on a set of subreddits that were publicly listed on Reddit's web interface in November 2024 (and July 2023).
Our sample is thus missing subreddits that are public but unlisted, which makes it likely that certain types of subreddit are underrepresented, such as NSFW subreddits. 
Our sample also excluded subreddits with a primary language other than English. 
Future research could explore whether the trends that we observed still hold across the types of communities excluded from our sample.
Our study also limits our analysis to text that is contained within a subreddit's ``rules'' metadata field. 
There are various other ways that online communities  communicate rules and norms, such as through public discussions, wikis, direct messages, or moderation actions. 
Future work interested in studying communities' attitudes towards AI use could consider these other sources of data, as well as qualitative approaches that would elicit direct input from community members and moderators.

\minititle{Ethics Statement.}
Our study was not required to undergo IRB review as we worked only with publicly available data.
Still, we recognize the importance of conducting ethical social media research and have taken steps to follow the best practices of the HCI community~\cite{fieslerRememberHumanSystematic2024}.
We did not collect any information about individual users or conversations, only community-level metadata.
Additionally, we did not include in our study any subreddits who were not listed in Reddit's public index of communities.
We make our datasets public with the publication of this work.

\section{Conclusion}
\balance
Communities across Reddit have responded to the arrival of generative AI by updating their public community rules with explicit positions on the use of this new technology.
A specific community's attitude towards the use of AIGC is nuanced and context-dependent, but our findings suggest a general trend of opposition, especially from content generation communities that value quality and authenticity.
It remains to be seen how these rules will be enforced, but HCI researchers should stay attuned to the changing needs and norms of these online communities as they continue to adapt to this new technology.
Especially as social platforms begin incorporating generative AI features, it is important for designers to consider how to encourage the self-determination of communities with nuanced and varied stances on the use of AIGC.

\begin{acks}
This material is based upon work supported by the National Science Foundation CISE Graduate Fellowships under Grant No. 2313998, as well as Grant No. CHS 1901151/1901329.
Any opinions, findings, and conclusions or recommendations expressed in this material are those of the authors and do not necessarily reflect the views of the National Science Foundation.
\end{acks}

\bibliographystyle{ACM-Reference-Format}
\bibliography{paper}

\newpage
\onecolumn
\appendix

\section{Appendix: LLM Prompts}
\subsection{Detecting Community Rules Governing AI Use}\label{app:prompt-detect}
\begin{lstlisting}
You will be given the public community rules for a subreddit. You will determine if any of the subreddit`s rules explicitly govern the use of AI tools or the content that such tools produce. Please return a JSON object with your determination, either YES or NO, and your explanation.

Examples:

1. Input: - AI generated(AI generated ): No AI (chatGPT etc.) generated messages/comments. No questions about chatGPT/AI generated code.
Output: {"determination": "YES", "explanation":"This rule bans code generated with ChatGPT or other AI tools"}

2. Input: - No bots(No bots): No bots. All bots are banned unless the bot provides exceptional value to /r/ MealPrepSunday.
Output: {"determination": "NO", "explanation":"This rule bans bots, not AI. Bots are not necessarily AI."}

3. Input: - Tributes, Split Edit, Deepfakes, roleplay, and DM Request are not allowed(Tributes, Split Edit, Deepfakes, roleplay, and DM Request are not allowed): The mods will take strict action against members who didn't follow this rule. The users who post Tributes will suffer a temporary ban.
Output: {"determination": "YES", "explanation":"This rule states that Deepfakes, which are a form of AI-generated content, are not allowed."}

4. Input: Rule 2 - Future Focus(Rule 2 - Future Focus): Submissions must be *future focused*. All posts must have an initial comment, a **Submission Statement**, that suggests a line of future-focused discussion for the topic posted. We want this submission statement to elaborate on the topic being posted and suggest how it might be discussed in relation to the future.
AI-focused posts are only allowed on the weekend.
Output: {"determination": "NO", "explanation":"This rule governs the use of posts about AI, not those made using it."}

5. Input: - Restricted accounts(New, Low Karma, bot, and underage accounts are restricted): * New or low karma accounts - may not post, can participate in the comments
* Negative karma accounts  - may not post or comment
* Under 18 users - may not post or seek diet advice
* No bot / AI accounts
Output: {"determination": "YES", "explanation":"This rule bans AI accounts."}

6. Input: - Rule 1: Must be relevant to language models(Rule 1: Must be relevant to language models): All language models are welcome here. 

Dall-E, GPT-3, GPT-Neo, GPT-J, ...
Output: {"determination": "NO", "explanation":"This rule says that posts must be about AI but says nothing about their use."}

7. Input: - No fakes or overly shopped photos, images must be posted as published. No memes.(No fakes or overly shopped photos, images must be posted as published. No memes.): NO FAKES OR OVERLY SHOPPED PHOTOS - may result in a ban, serious offense. Images must be posted as published except for Quality AI-enhanced images. We ask that those titles include ""Enhanced."" Cartoonish/amateurish attempts will be removed solely at the mod's discretion. Don't complain. Do better. ! No X-rays or ""bubbling."" No watermarks. No emojis. No weird borders on the images (i.e., white or black bars). No text on images.
Output: {"determination": "YES", "explanation":"This rule specifies that all AI-enhanced images must be titled 'Enhanced'"}

8. Input: Properly Source Fan Content or Concept Art(This fan/concept art isn't properly sourced): We like to support the community and creators of the Marvel Cinematic Universe as much as possible and properly crediting them for their work is the least we can do. When submitting fan art or a piece of concept art, we ask that you properly link to the original source and mention the name of the artist in the title.
Output: {"determination": "NO", "explanation":"This rule does not mention art produced by AI."}
\end{lstlisting}

\newpage
\subsection{Applying Rule Classification Labels}\label{app:prompt-rulelabels}
\begin{lstlisting}
You will be given one of a subreddit's public community rules and must determine which labels apply.
Follow the instructions for each label and determine if the label applies to the rule.
Return a JSON object where the keys are the label names and the values are TRUE if the label applies and FALSE if not.

Category: rule types

Label name: Advertising and Commercialization
Instructions: Is this rule about advertising and commercialization in the community?

Label name: Consequences/Moderation/Enforcement
Instructions: Does this rule specify what will happen if the rule is broken?

Label name: Content/Behavior
Instructions: Does this rule govern the types of behavior or content that are allowed or prohibited in the subreddit?

Label name: Copyright/Piracy
Instructions: Does this rule have to do with copyright, piracy, or giving credit to content's original creator?

Label name: Doxxing/Personal Info
Instructions: Is this rule about doxxing or revealing someone's personal information?

Label name: Format
Instructions: Does this rule specify that posts must follow a specific format?

Label name: Harassment
Instructions: Is this rule about harassment?

Label name: Hate Speech
Instructions: Is this rule about hate speech?

Label name: Images
Instructions: Is this rule about posting visual content like images, photos, or paintings?

Label name: Links and Outside Content
Instructions: Is this rule about posting links or outside content?

Label name: Low-Quality Content
Instructions: Is this rule about posting low-quality content?

Label name: NSFW
Instructions: Is this rule about posting NSFW content?

Label name: Off-topic
Instructions: Is this rule about posting off-topic content?

Label name: Personal Army
Instructions: Is this rule about using a group of other members to argue on your behalf?

Label name: Personality
Instructions: Does this rule govern what personality traits belong in the community?

Label name: Politics
Instructions: Does this rule govern posts about politics?

Label name: Prescriptive
Instructions: Does this rule include any positive statements about what users must do, or what is allowed?

Label name: Reddiquette/Sitewide
Instructions: Does this rule reference sitewide policy or norms?

Label name: Reposting
Instructions: Is this rule about reposting content that has already been posted?

Label name: Restrictive
Instructions: Does this rule include any negative statements about what users must not, or are not allowed to, do?

Label name: Spam
Instructions: Is this rule about posting spam, such as karma farming, or posting a large volume of content?

Label name: Spoilers
Instructions: Is this rule about posting spoilers?

Label name: Trolling
Instructions: Is this rule about trolling?

Label name: Voting
Instructions: Is this rule about voting in the community?

Category: Usage
Label name: usage--generate
Instructions: Does this rule use variations of the words "create" or "generate" to discuss content made with AI, such as AI-generated content?

Label name: usage--assist
Instructions: Does this rule reference using AI assistance, for example to enhance, alter, or edit images?

Label name: usage--general
Instructions: Does this rule mention AI, but does not mention a specific type of use?

Category: Rationale
Label name: rationale--lowquality
Instructions: Does this rule explicitly mention that content made with AI is low quality?

Label name: rationale--loweffort
Instructions: Does this rule mention that content made with AI is low effort?

Label name: rationale--spam
Instructions: Does this rule mention that content made with AI is spam?

Label name: rationale--misleading
Instructions: Does this rule mention that content made with AI is misleading?

Label name: rationale--inaccurate
Instructions: Does this rule mention that content made with AI is inaccurate?

Label name: rationale--inauthentic
Instructions: Does this rule mention that content made with AI is fake or inauthentic?

Label name: rationale--inhuman
Instructions: Does this rule mention that content made with AI is inhuman?

Label name: rationale--offtopic
Instructions: Does this rule mention that content made with AI is off topic?

Label name: rationale--policy
Instructions: Does this rule mention laws or site policies?

Label name: rationale--ethics
Instructions: Does this rule mention an ethical or moral objection to AI?

Label name: rationale--original_content
Instructions: Does this rule mention that content made with AI is not original content?

Category: Medium
Label name: form--text
Instructions: Does this rule explicitly mention text made with AI?

Label name: form--images
Instructions: Does this rule explicitly mention images generated or enhanced by AI?

Label name: form--videos
Instructions: Does this rule explicitly mention videos made with AI?

Label name: form--art
Instructions: Does this rule explicitly mention art made with AI?

Label name: form--deepfake
Instructions: Does this rule explicitly mention deepfakes?

Label name: form--fakes
Instructions: Does this rule explicitly mention fakes?

Label name: form--bots
Instructions: Does this rule explicitly mention AI bots?

Label name: form--content
Instructions: Does this rule explicitly mention AI content, for example AI-generated content?

Label name: form--general
Instructions: Does this rule govern the use of AI without explicitly referencing a specific type of use?

Category: Stance
Label name: stance--unqualified_ban
Instructions: Does this rule ban or disallow AI or some sort of content made using AI without providing exceptions?

Label name: stance--qualified_ban
Instructions: Does this rule ban some forms of AI use, but allow it in certain circumstances or if users label it?

Label name: stance--disclosure
Instructions: Does this rule require that users identify when the content that they post is made to some extent using AI, such as with a label, or flair, or certain text in the title?

Label name: stance--enabling
Instructions: Does this rule explictly say that some content made with AI is allowed under certain circumstances, for example if the fact that AI was used is disclosed?

Label name: stance--requiring
Instructions: Does this rule explictly require that posted content must be made with AI?
\end{lstlisting}

\subsection{Applying Subreddit Classification Labels}\label{app:prompt-sublabels}
\begin{lstlisting}
You are to assist with a task to qualitatively code Reddit communities, which are known as subreddits.
You will be given a JSON string containing metadata about a subreddit that will contains its name, public description, description and rules.
For each category below, follow the instructions to determine which labels apply to the subreddit.
Return a JSON object where they keys are "Topical Question and Answer", "Learning and Perspective Broadening", "Social Support", "Content Generation", "Affiliation with an Entity", "Topic" and the values are the most appropriate label values.

Category: Community Archetype
Instructions: An archetype describes an abstract set of qualities that are likely to co-occur in individual instances of an object. A Community Archetype applies this concept to subreddits and classifies them into types of communities based on identifiable subreddit features and common user behaviors. The following labels each correspond to a Community Archetype. Consider each label and the characteristics associated with it and determine if it describes the given subreddit. For each label, choose the value Y if it best describes the subreddit and N if it does not.
Labels:
- Name: Topical Question and Answer
  Characteristics: Subreddits that are explicitly set up with strict rules to create a Question and Answer format around specific topics. Most top-level posts are questions, while most comments are answers, or discussions and enrichments of others` answers.
- Name: Learning and Perspective Broadening
  Characteristics: Subreddits that users visit primarily to learn or broaden their perspective. Top-level posts are often pointers to current events, publications, or relevant news, or relatable personal stories, experiences, and questions. Comments tend to elaborate upon the ideas raised, for example by celebrating or disagreeing, making jokes, providing contrasting or similar personal stories and experiences, or adding new ideas and references into the mix.
- Name: Social Support
  Characteristics: Subreddits where socially supportive behaviors are the main purpose and organizing principle for gathering. Top-level posts are often specific personal experiences or sensitive disclosures, questions about a health issue, announcements of milestones, reflections or venting, or sharing of resources, artwork, and encouraging thoughts and memes/jokes. Comments generally offer support, reflection, commiseration, resources, and validation.
- Name: Content Generation
  Characteristics: Subreddits that are devoted to content with a specific format, sense of humor, viewpoint, purpose, or type of creative expression. Top-level posts exemplify a subreddit`s specific content style and comments often include peoples` opinions on the content, extra information about the content, or sometimes commiseration with the original poster.
- Affiliation with an Entity
  Characteristics: Subreddits that have explicit affiliations with particular entities, such as geographical places or organizations (cities, universities), popular media (book or fan series, TV shows), sports teams, etc. Posts often feature local or breaking news and events related to the entity. In the comments, users generally express their feelings about news or events, and answer questions or offer advice and opinions about the entity.

Category: Topic
Instructions: Pick one of the following labels that best describes the topic of the subreddit. If multiple labels apply, choose the one that is most descriptive.
Labels:
- Video Games
- Image Sharing
- Entertainment
- Personal
- Technology and Science
- Celebrity
- Hobby
- Local
- Sports
- Support
- Music
- Art
- Humor
- Meta
- Business and Organizations
- Animals
- Work
- Writing
- Learning
- Politics
- Games
- News
- Health
- Food
- Parody
- Culture
- Fashion
- Drugs
\end{lstlisting}
\end{document}